\documentclass{article}
\usepackage{setspace}
\usepackage[utf8]{inputenc}
\usepackage{amsmath}
\usepackage{slashed}
\usepackage{amssymb}
\usepackage{bbold}
\usepackage{halloweenmath}
\usepackage{graphicx}
\usepackage{authblk}
\usepackage{caption}
\graphicspath{ {./images/} }
\usepackage{comment}
\usepackage{tikz}
\usepackage{geometry}
\usepackage{hyperref}
\hypersetup{ 
colorlinks=true,
linkcolor=blue,
filecolor=magenta,
urlcolor=blue,
citecolor=blue,
pdfpagemode=FullScreen
}
\usepackage[numbers,sort&compress]{natbib}

\usepackage[normalem]{ulem}
\usepackage{enumerate}
\usepackage[shortlabels]{enumitem}

\title{Ghost Hunting in the Yang-Mills Vacuum}
\author[1]{Seth Grable}

\affil[1]{Department of Physics, University of Colorado Boulder, Colorado 80309, USA}
\date{July, 2026}

\begin{document}
\begin{spacing}{1.5}
\maketitle

\begin{abstract}
In this work, I analyze the zero modes of a one-loop semiclassical Yang-Mills theory in \(3+1\)d. I find that zero modes are generated by gauge redundancy of the background field. Proper gauge fixing, achieved by introducing a bosonic ghost term, together with zeta-function regularization, yields a finite one-loop effective action in closed form that reproduces the well-known one-loop Yang-Mills beta function.

\end{abstract}

\section{Introduction}\label{sec 1}
\paragraph{}
In this work, I will consider a semiclassical expansion of the Yang-Mills vacuum. A semiclassical expansion of the action around some field configuration \(\bar{\phi}\) is one that satisfies the classical equations of motion plus small fluctuations around the classical field configuration. For a field \(\phi= \bar{\phi} + \eta\) where \(\bar{\phi}\) satisfies the classical equations of motion and \(\eta\) are fluctuations around the classical configuration, a variation of the action gives
\begin{equation}
    S[\bar{\phi}+\eta] =
   S[\bar{\phi}]  + \frac{1}{2} \eta\frac{\delta^2 S[\phi]}{\delta \phi^2}\eta\bigg|_{\bar{\phi}}  + \mathcal{O}(\eta^3).
\end{equation}
 For a Euclidean path integral \(Z=\int d\phi \mathcal{D}\eta e^{-S/\hbar}\) under the substitution \(\eta\rightarrow\sqrt{\hbar}\eta\), the expansion is valid for \(\hbar\ll 1\). Equivalently, in natural units with \(\hbar = 1\), this is an expansion in small fluctuations \(\eta\) around the classical action, analogous to a WKB expansion. If \(\bar{\phi}\) represents a true stable minimum of the theory, then the variational principle guarantees \(S[\bar{\phi}]<S[\bar{\phi}+\eta]\). Then in the Euclidean metric this results in a classically dominated path integral with exponential suppression of the fluctuations as \(e^{-S[\bar{\phi}]}> e^{-S[\bar{\phi}+ \eta]}\), giving the dominant contributions of \(Z\) as
 \begin{equation}\label{semi class}
     Z \approx \int\mathcal{D}\eta \exp{\Bigg[-\Bigg( S[\bar{\phi}]  + \frac{1}{2} \eta\frac{\delta^2 S[\phi]}{\delta \phi^2}\eta\bigg|_{\bar{\phi}} \dots\Bigg)\Bigg]}
 \end{equation}
The dominance of the classical trajectory in the quantum amplitude was first identified by Dirac \cite{dirac2005lagrangian}, who showed the transition amplitude is proportional to \(e^{iS_{cl}/\hbar}\) for small \(\hbar\). However, a stronger claim can be made by viewing \(Z\) as the Euclidean metric version of the Feynman sum as \(Z=\sum_{\text{paths}}e^{-S}\), giving a suppression of large \(S\). Cubic and quartic order variations of the action will give small contributions from paths (or field configurations) close to the classical configuration, which minimizes \(S\). Similarly, higher order variations will give large contributions from paths that are far away from the classical configuration giving overall large values of \(S\). These small contributions which are relevant for the calculation of \(Z\) are approximated as zero in \eqref{semi class}, rendering the approximation valid only for a strictly positive second variation of the action.  
 
Integrating out all fluctuations at quadratic order is equivalent to summing over all one-loop contributions. In this sense, the classical expansion generates non-perturbative physics not accessible by standard perturbation theory \cite{grable2023fully,grable2022theremal, romatschke2019finite}. 

 Semiclassical Yang-Mills theories date back to the late 1970s and have been studied by Savvidy, Leutwyler, Nielsen, Olesen, and others \cite{savvidy1977infrared, leutwyler1981constant, nielsen1978approximate, nielsen1978unstable, grable2025vanishing}.
 Savvidy originally noted that the semiclassical theory is unstable due to negative eigenvalues in the one-loop operator \cite{savvidy1977infrared}. However, Savvidy has recently proposed \cite{savvidy2023stability} a self-dual background configuration which renders the one-loop operator positive semidefinite, leaving zero modes as the remaining obstacle.

\section{The Background Field Set up for Yang-Mills}\label{Sec 2}
\paragraph{}
A Yang-Mills Lagrangian density for general SU(N) is given by
\begin{equation}\label{L}
    L=\frac{1}{4}(\mathcal{F}_{\mu\nu}^a)^2.
\end{equation}
 The adjoint representation defines the field strength tensor and covariant derivative as:
\begin{equation}\label{eq 2}
\begin{split}
   & \mathcal{F}_{\mu\nu}^a = \partial_\mu\delta^{ac} \mathcal{A}^c_\nu(x) -\partial_\nu\delta^{ac} \mathcal{A}^c_\mu(x) + g_0f^{abc}\mathcal{A}^b_\mu(x)\mathcal{A}^c_\nu(x)\\&
   D^{ac}_\mu = \partial_\mu\delta^{ac} + g_0f^{abc}\mathcal{A}^b_\mu(x),
\end{split}    
\end{equation}
where \(f^{abc}\) are the structure constants of an \(SU(N)\) algebra, and \(g_0\) is the bare Yang-Mills coupling constant. In Euclidean space with the temporal direction compactified on the thermal cylinder, the partition function is \cite{laine2016basics}
\begin{equation}\label{Z}
    Z=\int\mathcal{D}\mathcal{A} e^{-\frac{1}{4} \int_x (\mathcal{F}_{\mu\nu}^a)^2}.
\end{equation}
 The gauge fields can be separated into a background field \(A^a_\mu(x)\) and fluctuations \(a^a_{\mu}(x)\) with the change of variables \(\mathcal{A}^a_\mu(x)=\frac{A^a_\mu(x)}{g_0}+a^a_\mu(x)\). This expands the field strength tensor to \cite{peskin2018introduction,weinberg1995quantum}:
\begin{equation}\label{F}
\begin{split}
   & \mathcal{F}_{\mu\nu}^a = \frac{1}{g_0}F_{\mu\nu}^a + D^{ac}_\mu\ a^c_\nu(x) -D^{ac}_\nu\  a^c_\mu(x) + g_0f^{abc}a^b_\mu(x)a^c_\nu(x)\\&
   D^{ac}_\mu = \partial_\mu\delta^{ac} + f^{abc}A^b_\mu(x)\\&
   F_{\mu\nu}^a= \partial_\mu\delta^{ac}A^c_\nu(x) -\partial_\nu\delta^{ac}A^c_\mu(x) +f^{abc}A^b_\mu(x)A^c_\nu(x).
\end{split}    
\end{equation}
Next, the background field \(A^a_\mu\) is chosen as a classical self-dual and source-free configuration, meaning 
\begin{equation}\label{Const F}
    F^a_{\mu\nu}=\begin{pmatrix}
        0&-B^a&0&0\\B^a&0&0&0\\0&0&0&-B^a\\0&0&B^a&0
    \end{pmatrix},
\end{equation} 
and
\begin{equation}\label{source free}
     D^{ab}_{\mu} F^b_{\mu\nu} =0.
 \end{equation}
 This configuration is given by
\begin{equation}\label{A}
     A^a_\mu(x) = -\frac{1}{2}F^a_{\mu\nu}x_\nu,
 \end{equation}
 The structures of \(F_{\mu\nu}^a\) in equations \eqref{F} and \eqref{Const F} are consistent with the substitution of \(  A^a_\mu(x) = -\frac{1}{2}F^a_{\mu\nu}x_\nu\) as
 \begin{equation}\label{B id}
     \frac{1}{4}f^{abc}F^a_{\mu\nu}x_\nu F^c_{\nu\mu}x_\mu =0
 \end{equation}
due to the symmetric configuration of gauge fields and the anti-symmetry of the structure constants. To see that the choice of background field in \eqref{A} satisfies
 \begin{equation}\label{source free}
     D^{ab}_{\mu} F^b_{\mu\nu} =0,
 \end{equation}
note the identity in equation \eqref{B id}; then \eqref{source free} reduces to spatial derivatives acting on a constant field strength tensor. In addition, I make the choice that \(B^a\) is a constant vector in color space, writing \(B^a = B e^a\), where \(e^a\) are components of a unit vector in color space. This is designed such that the chromomagnetic fields contract with \(T^a_{bc}\), the generators of \(SU(N)\) in the adjoint representation, to give 
\begin{equation}
    B^a T^a_{bc}= B e^aT^a_{bc}
\end{equation} where \(B\) is a constant with dimensions of mass-squared. Additionally, defining
\begin{equation}\label{C}
    e^aT^a_{bc} = C^{bc},
\end{equation}
I work in a diagonal color-space basis where 
\begin{equation}\label{d}
    \mathbf{d}=\mathbf{U}^{-1}\mathbf{C}\mathbf{U}
\end{equation}
where \(\mathbf{U}\) is a unitary operator in color space, and \(d^{cc}\) are the eigenvalues of the diagonal matrix \(\mathbf d\), which has a zero trace. At this stage, no approximations have been made — the resulting Lagrangian is an exact rewriting of equation \eqref{L} under a change of variables and a global unitary transformation on color space.

The self-dual and covariantly constant background configuration has several key features: it satisfies the classical source-free Yang-Mills equations of motion; it reproduces the known Yang-Mills \(\beta\)-function \cite{peskin2018introduction}; and the quadratic operator in the path integral has no negative eigenvalues. Negative modes of the quadratic operator under non-self-dual background configurations generate what is often referred to as a Nielsen-Olsen instability \cite{nielsen1978unstable}. This instability is removed by the self-dual configuration, and what remains is to understand the zero modes of the one-loop effective action. 

\section{Gauge Fixing}
\paragraph{}
A standard gauge fixing condition is \(D^{ac}_\mu a^c_\mu(x) =0 \) \cite{t1976computation,savvidy1977infrared}, giving the gauge-fixed Lagrangian as
\begin{equation}\label{eq a}
\begin{split}
   & \mathcal{L} = \frac{1}{4}\Big[\Big(\frac{1}{g_0}F^a_{\mu\nu} + D^{ac}_{\mu}a^c_\nu - D^{ac}_{\nu}a^c_\mu + g_0f^{abc}a_\mu^b a_\nu^c \Big)^2  + 2(D^{ac}_\mu a^c_\mu)^2 \Big] +\bar{c}^a[(D^2)^{ac}]c^c.
\end{split}   
\end{equation}
where I have used an analog of the Feynman-'t Hooft gauge in the Gaussian contribution of the gauge fixing procedure. Equation \eqref{eq a} is invariant under local gauge transformations of \cite{peskin2018introduction}
\begin{equation}\label{transforms}
\begin{split}
        &a^a_\mu(x) \rightarrow a^a_\mu(x)- f^{abc}\beta^b(x) a^c_\mu(x)\\&
        A^a_\mu(x) \rightarrow A^a_\mu(x) + D_\mu \beta^a(x) \\&
        c^a(x)\rightarrow c^a(x) -f^{abc}\beta^b(x) c^c(x).
\end{split}
\end{equation}
Integration by parts can be performed over the gauge fluctuations \(a^a_\nu\) and equation \eqref{Z} becomes
\begin{equation}\label{eq 12}
    Z=\int dA\int \mathcal{D}\bar{c}\mathcal{D}c\mathcal{D}a e^{-\int_x \frac{1}{4g^2_0}(F^a_{\mu\nu})^2-S_0-S_I}
\end{equation}
with \cite{grable2023elements,savvidy2023stability}
\begin{equation}\label{}
    \begin{split}\label{eq 13}
        &S_0 = \int_x \frac{1}{2}a^a_\mu\Big[-(D^2)^{ac}\delta_{\mu\nu} - 2F_{\mu\nu}^bf^{abc}\Big]a^c_\nu + \bar{c}^a(-D^2)^{ac}c^c \\&
       S_I=\int_x g_0(D_\mu a_\nu^a)f^{abc} a_\mu^b a_\nu^c + \frac{g_0^2}{4}(f^{abc}a_\mu^b a_\nu^c)^2. 
    \end{split}
\end{equation}
The operators in quadratic terms are relabeled as \(\hat{\theta}_{\text{Glue}}\equiv-(D^2)^{ac}\delta_{\mu\nu} - 2F_{\mu\nu}^bf^{abc}\) and \(\hat{\theta}_{\text{Ghost}}\equiv-(D^2)^{ac} \) presenting the one-loop effective theory as \cite{peskin2018introduction}
\begin{equation}\label{one-loop}
  Z\approx \int dA e^{-\frac{\beta V}{4g^2_0}(F^a_{\mu\nu})^2-\frac{1}{2}\ln\det(\Hat{\theta}_{\text{Glue}}) +\ln\det(\Hat{\theta}_{\text{Ghost}})}.
\end{equation}
 The quadratic gauge and ghost field operators transform covariantly under Lorentz boosts. However, the spectra of \(\hat{\theta}_{\text{Glue}}\) and \(\hat{\theta}_{\text{Ghost}}\) are scalar quantities, as shown in Section \ref{Sec 3}. Thus, \eqref{one-loop} is Lorentz invariant. The gauge-fixed Lagrangian is BRST invariant under a standard quantization procedure \cite{peskin2018introduction}, implemented with the background gauge condition of \(D^{ac}_\mu a^c_\mu(x) = 0\). Setting \(g\to 0\) generates the one-loop BRST and Lorentz invariant effective action. Although we have gauge fixed and truncated \(S\) in terms of \(a^a_\mu\), since all functions of \(A^a_\mu(x)\) in \(S_0\) are multiples of \(D^{ac}_\mu\) and \(F^a_{\mu\nu}\), \(S_0\) remains invariant under local transformations of the background field \(A^a_\mu(x)\).
The residual background field gauge invariance is often noted as a defining feature of the theory \cite{abbott1981background}.
However, generally a lack of gauge fixing generates zero modes from integrating over contributions where the action is flat in the direction of the non-physical gauge transformation. In fact, the gauge orbit of the residual background symmetry generates zero modes in the one-loop effective action. This can be shown by expanding the action around the background field as \(A^a_\mu = B^a_\mu + a^a_\mu\). In the \(g_0\to 0\) limit, and where \(A^a_\mu\) satisfies the classical equations of motion, this gives
\begin{equation}\label{action fluc}
    S[B^a_\mu + a^a_\mu] = S[B^a_\mu]  + \frac{1}{2} a^a_\mu\frac{\delta^2 S[A^a_\mu]}{\delta A^2}\bigg|_{B^a_\mu} a^c_\nu.
\end{equation} 
Consider the gauge transformed action with respect to the background field, by gauge invariance we must have 
\begin{equation}\label{gauge invariance}
    S[B^a_\mu + D_\mu \beta^a(x)] = S[B^a_\mu].
\end{equation}
Thus, in the \(g_0 \to 0\) limit zero modes of the quadratic operator are generated by gauge invariance of the classical background. Now, letting \(g_0\) be nonzero, by gauge invariance, the cubic and quartic terms sum to zero as the quadratic contribution is independent of \(g_0\).
Thus, the one-loop zero modes of the gauge fluctuations are synonymous with zero modes generated by the gauge invariance of the classical background. The zero mode divergence is not resolved in the gauge fixing procedure for \(a^a_\mu(x)\). Further, because the ghost determinant depends on the background field, removing zero modes introduces surplus ghost degrees of freedom. These surplus ghosts correspond to the eigenvalues of \(-D^{2}\) acting on \(\omega^a_\mu = D_\mu \beta^a(x)\), where \(\omega^a_\mu\) is along the gauge orbit of \(A_{\mu }^{a}\) as defined in Eq. \eqref{transforms}.

To gauge fix the background field or, likewise, to remove the surplus fermionic ghost terms, let \(D^a_\mu {\omega}^a_\mu = 0\) by applying \(\delta\big(D^a_\mu {\omega}^a_\mu \big)\) to the one-loop theory. Where the subscript \(\omega^a_\mu\) in the following indicates that the determinant is only taken as the operator acting on \(\omega^a_\mu\), this gives 
\begin{equation}\label{B gauege fix}
    \int d \omega \delta\big( D_\mu \omega^a_\mu\big) e^{-\omega^a_\mu [\hat{\theta}]\omega^a_\mu} = \frac{1}{\det(D^a_\mu)\big|_{\omega^a_\mu}},
\end{equation}
and as \(D^a_\mu\) is anti-Hermitian, \(\det(D^a_\mu) = \sqrt{\det\big(-D^2\big)}\). This scalar ghost term, referred to as a Nielsen-Kallosh ghost \cite{nielsen1981brs, kallosh1978modified}, is shown to cancel surplus fermionic ghost terms in section \ref{Sec 4}, yielding the correct \(\beta\)-function. Removing the zero modes and introducing the bosonic ghost term given in \eqref{B gauege fix} removes one BRST doublet, leaving the remaining theory BRST invariant.

\section{The One-Loop Eigenspectrum}\label{Sec 3}
\paragraph{}
To find the eigenspectrum of \(\theta_{\text{Glue}}\), the covariant derivative and the field strength tensor are considered separately. Because \(-(D^2)^{ac}\delta_{\mu\nu}\) and \(2F_{\mu\nu}^bf^{abc}\) are simultaneously diagonalizable in Lorentz and color space \cite{yildiz1980vacuum,grable2023elements} the sum of the eigenvalues of \(-(D^2)^{ac}\delta_{\mu\nu}\) and \(2F_{\mu\nu}^bf^{abc}\) equals the eigenvalues of \(-(D^2)^{ac}\delta_{\mu\nu}+2F_{\mu\nu}^bf^{abc}\). Equations \eqref{A} and \eqref{F} give the covariant derivative in terms of the field strength tensor as
\begin{equation}
    D^{ac}_\mu  = \partial_\mu \delta^{ac} +\frac{i}{2}C^{ac}F_{\mu\nu}x_\nu
\end{equation}
where \(C^{ac}\) defined in \eqref{C}. Letting the Lorentz index run from 0 to 3 gives \cite{savvidy2023stability}
\begin{equation}\label{Big D}
\begin{split}
        -(D^2)^{ac} =  &-(\partial^2_0)\delta^{ac}  -(\partial^2_1)\delta^{ac}  + i(BC)^{ac}\big(x_1\partial_0-x_0\partial_1\big) + \frac{1}{4}(B^2C^2)^{ac}\big(x_0^2+x_1^2 \big)\\&
    -(\partial^2_2)\delta^{ac}  -(\partial^2_3)\delta^{ac}  + i(BC)^{ac}\big(x_3\partial_2-x_2\partial_3\big) + \frac{1}{4}(B^2C^2)^{ac}\big(x_2^2+x_3^2\big).
\end{split}       
\end{equation}
As \(d^{cc}\), the eigenvalues of the color matrix \(\mathbf{d}\), defined is \eqref{d} come in plus/minus pairs, there are two forms of \eqref{Big D} to consider. For example, in the \(0-1\) plane in Lorentz space, where let \(d^{aa} = \lambda\), the two distinct cases are:
Case 1), for positive \(\lambda\)
\begin{equation}
  -D^2_{+}(\lambda) =  -\partial^2_0  -\partial^2_1  + iB|\lambda|\big(x_1\partial_0-x_0\partial_1\big) + \frac{1}{4}(B^2\lambda^2)\big(x_0^2+x_1^2 \big),
\end{equation}
and Case 2) for negative \(\lambda\)
\begin{equation}
     -D^2_-(\lambda) =  -\partial^2_0  -\partial^2_1  - i B|\lambda|\big(x_1\partial_0-x_0\partial_1\big) + \frac{1}{4}(B^2\lambda^2)\big(x_0^2+x_1^2 \big).
\end{equation}
Generalizing back to all 4 spatial dimensions, for Case 1) I define the creation and annihilation operators as 
\begin{equation}\label{operators}
\begin{split}
    & c_\mu =\Bigg[ \partial_\mu + \frac{1}{2}B|\lambda|x_\mu\Bigg], \quad  c^{\dagger}_\mu = \Bigg[-\partial_\mu + \frac{1}{2}B|\lambda| x_\mu\Bigg],
\end{split}
\end{equation}
 Using the canonical commutation relation \([x_\mu, -i\partial_\nu]=i\delta_{\mu\nu}\) it can be shown that \([c_\mu,c^{\dagger}_\mu]=B|\lambda|\). Letting \((b^\dagger_{\mu\nu})_+ = c^{\dagger}_\mu+ic^{\dagger}_\nu\) and \((b_{\mu\nu})_+ = c_\mu-ic_\nu\)  \cite{leutwyler1981constant} direct calculation gives
\begin{equation}
\begin{split}\label{eq opps}
     -D^2_+(\lambda) = (b^{\dagger}_{01})_+ (b_{01})_+ + (b^{\dagger}_{23})_+ (b_{23})_+ + 2B|\lambda|.
\end{split}
\end{equation}
The spectrum of \(-D^2_+(\lambda)\) is defined by the commutator \([(b_{\mu\nu})_+, (b^{\dagger}_{\mu\nu})_+] = B|\lambda|\).
Similarly for Case 2) we can define \((b^\dagger_{\mu\nu})_- = c^{\dagger}_\mu-ic^{\dagger}_\nu\) and \((b_{\mu\nu})_- = c_\mu+ic_\nu\). This gives
\begin{equation}
      -D^2_-(\lambda) = (b^{\dagger}_{01})_- (b_{01})_- + (b^{\dagger}_{23})_- (b_{23})_- + 2B|\lambda|.
\end{equation}
Similarly the spectrum of \(-D^2_-(\lambda)\) is defined by the commutator \([(b_{\mu\nu})_-, (b^{\dagger}_{\mu\nu})_-] = B|\lambda|\), and a similar analysis can be done for positive and negative values of \(B\) giving the full spectrum of \(-D^2(\lambda)\) as 
\begin{equation}\label{eq 23}
    \Omega= \sum_{m,n} |B\textbf{d}|\Big((2n+1) + (2m+1)\Big),
\end{equation}
drawing consistency with the positive and real eigenspectrum of the canonical quantum harmonic oscillator. Finally, the eigenvalues of \(F^{ac}_{\mu\nu}\) come in complex pairs of \(\pm i B \lambda\) giving the total eigenspectrum of \(\theta_{\text{Glue}}\) as ,
\begin{equation}\label{eq 20}
\begin{split}
    &\Lambda(\lambda)^+_{m,n} =|B\lambda|\Big( (2n+1) +(2m+1) +2\Big) \\&
    \Lambda(\lambda)^-_{m,n} = |B\lambda|\Big((2n+1)+(2m+1)-2\Big),
\end{split}
\end{equation}
with \(m,n\in \mathbb{N}_0\). The spectrum of  \(\Lambda(\lambda)^-_{m,n}\) contains a zero eigenvalue where \(m=n=0\). This mode is associated with the gauge invariance of the background gauge \(A^a_\mu(x)\) shown in \eqref{action fluc} and \eqref{gauge invariance}. As the eigenfields of the quadratic theory are associated with harmonic oscillator states, the states themselves are not invariant with respect to Lorentz Boosts in the Minkowski metric or similarly translations plus global rotations in the Euclidean metric. However, since \(S_0\) remains a scalar, \(S_0\) and \(Z\) are invariant to translations and global rotations in the Euclidean metric. That is, the statistics or interactions of theory maintain the required invariance principles of an interacting field theory. The symmetry breaking at the level of individual states of the theory is referred to as \textit{state symmetry breaking} in \cite{grable2026scalar}. State symmetry breaking is a generalization of spontaneous symmetry breaking in which there was no original symmetry of the states to spontaneously break, but individual states break a symmetry of \(Z\). In the next section I will show how the proper gauge fixing of \(A^a_\mu(x)\) and zeta function regularization of \(Z\) leads to the well-known Yang-Mills \(\beta\)-function along with a finite renormalized vacuum pressure. 

\section{Partition Function Calculation}\label{Sec 4}
\paragraph{}
In this section, I calculate the zero-temperature and infinite-volume pressure as \(P=\frac{\ln Z}{\beta V}\). To  calculate \(\ln Z = \sum_i c_i\ln\det(\hat{\theta_i})\) I use the identity first noted by Hawking \cite{hawking1977zeta} of
\begin{equation}
    \ln\det(\hat{\theta}) = -\frac{d}{ds}\bigg[\sum_i\frac{1}{\lambda_i^s}\bigg]_{s=0}.
\end{equation}
Using the integral definition of the gamma function, it follows then that
\begin{equation}\label{gam}
    \ln\det(\hat{\theta}) = -\frac{d}{ds}\bigg[\frac{1}{\Gamma(s)} \int^\infty_0 d\tau \tau^{s-1}k(\hat{\theta})\bigg]_{s=0}
\end{equation} where \(k(\hat{\theta}) = \text{Tr}^{ac}_{\mu\nu}\text{Deg}\sum_{n,l}e^{-\tau \lambda_{n,l}}\) is the heat kernel of \(\hat{\theta}\), and \(\text{Deg}\) is the degeneracy of the \(n=l=0\) state \cite{bertlmann2000anomalies}. The degeneracy of the gauge fluctuations is Deg \( = \beta V\frac{(B\lambda_i)^2}{16\pi^2}\) \cite{savvidy2023stability}. With this, and the inclusion of a running mass scale \(\mu^2\), the heat kernel for \(\Hat{\theta}_{\text{Glue}}\) is 
\begin{equation}
    K(\theta_{\text{Glue}})  = \sum_j\beta V\frac{ (B\lambda_j)^2}{4\pi^2}\frac{\cosh(2B|\lambda_j|\tau/\mu^2)}{\sinh^2( B|\lambda_j|\tau/\mu^2)}.
\end{equation}
 The \(\sinh(\cdot)\) term is generated by the sum over n and l, and the \(\cosh(\cdot)\) term is generated by the trace over \(e^{-\tau F^{ac}_{\mu\nu}}\). Similarly, the fermionic ghost kernel is given by 
\begin{equation}
     K(\theta_{\text{Ghost}}) = \sum_j\beta V\frac{ (B\lambda_j)^2}{16\pi^2}\Bigg(\frac{1}{\sinh^2( B|\lambda_j|\tau/\mu^2) } \Bigg),
\end{equation}
Using the identity 
\begin{equation}
    \frac{1}{\sinh^2(\tau)} = 4\sum^\infty_{n=0} n e^{-2\tau n}
\end{equation}
The \(K(\hat{\theta}_{\text{Glue}})\) and \(K(\hat{\theta}_{\text{Ghost}})\) are expressed as 
\begin{equation}\label{K glue}
     K(\hat{\theta}_{\text{Glue}}) = \sum_{j}\beta V\frac{ (B\lambda_j)^2}{2\pi^2}\sum_{n=0}^\infty n\Big(e^{-2B|\lambda_j|\tau(n-1)/\mu^2} + e^{-2B|\lambda_j|\tau(n+1)/\mu^2}\Big)
\end{equation}
and 
\begin{equation}\label{K-ghost}
     K(\Hat{\theta}_{\text{Ghost}}) = \sum_{j}\beta V\frac{ (B\lambda_j)^2}{4\pi^2}\sum_{n=0}^\infty ne^{-2B|\lambda_j|\tau n/\mu^2} 
\end{equation}
To properly account for the degeneracy of the bosonic (background field) ghost term, I take the \(\frac{1}{2}\) of the degeneracy given \eqref{K glue}, as the zero modes account for two and not four degrees of freedom in Lorentz space. This gives
\begin{equation}\label{B-ghost}
K(\Hat{\theta}^{\text{Background}}_{\text{Ghost}}) =\sum_{j}\beta V\frac{(B\lambda_j)^2}{2\pi^2} e^{-2B|\lambda_j|\tau/\mu^2},
\end{equation}
which is a term proportional to the \(n=1\) term in \eqref{K-ghost}. This must be the case, as the \(n=1\) term in is the lowest mode of ghost operator, which is the mode that becomes excessive after the zero mode, also generated by the \(n=1\) term in \eqref{K glue}), is removed.
Using \eqref{gam} 
\begin{equation}\begin{split}\label{lns 1}
   & -\frac{1}{2}\ln\det{\hat{\theta}_{\text{Glue}}} + \ln\det{\hat{\theta}_{\text{Ghost}}} -\frac{1}{2}  \ln\det{\Hat{\theta}^{\text{Background}}_{\text{Ghost}}} = \\&
 \sum_j\frac{d}{ds}\Bigg[\frac{\mu^4\beta V}{16\pi^2}  \bigg(\frac{2B|\lambda_j|}{\mu^2}\bigg)^{2-s}\Bigg(\sum^\infty_{n=0} \frac{n}{(n-1)^s} +  \frac{n}{(n+1)^s} - \frac{n}{n^{s}} +1 \Bigg)\Bigg]_{s=0}.
\end{split}    
\end{equation}
%pg 2148 
 A partial fraction decomposition of the first two sums in \eqref{lns 1} gives
\begin{equation}\label{decomp}
    \frac{n}{(n-1)^s} + \frac{n}{(n+1)^s} = \frac{1}{(n-1)^{s-1}} + \frac{1}{(n-1)^s} + \frac{1}{(n+1)^{s-1}} - \frac{1}{(n+1)^s}.
\end{equation}
The zero mode is explicit as the \(n=1\) term in the first sum as \(0^{-s}\). Removing this term due to the gauge fixing condition of \eqref{B gauege fix} and applying \(\zeta(s,1) = \zeta(s)\) gives
\begin{equation}\begin{split}\label{lns}
   & -\frac{1}{2}\ln\left(\det{\hat{\theta}_{\text{Glue}}}\right)_{\text{regulated}} + \ln\det{\hat{\theta}_{\text{Ghost}}} +  \ln\det{\Hat{\theta}^{\text{Background}}_{\text{Ghost}}} = \\&
 \sum_j\frac{d}{ds}\Bigg[\frac{\mu^4\beta V}{16\pi^2}  \bigg(\frac{2B|\lambda_j|}{\mu^2}\bigg)^{2-s}\Big(\zeta(s-1) + \zeta(s) + \zeta(s-1,1) - \zeta(s,1) - \zeta(s-1) + 1 \Big)\Bigg]_{s=0}=\\&
 \sum_j\frac{d}{ds}\Bigg[\frac{\mu^4\beta V}{16\pi^2}  \bigg(\frac{2B|\lambda_j|}{\mu^2}\bigg)^{2-s}\Big(\zeta(s-1) + 1 \Big)\Bigg]_{s=0}.
\end{split}    
\end{equation}
Incidentally, using the Hankel contour definition of the \(\zeta\) function, the zero modes can likewise be regulated \footnote{Removing zero modes with only the use of analytic continuation leaves surplus ghost degrees of freedom yielding an incorrect \(\beta\)-function, and a violation of BRST invariance. If you evaluate equation \eqref{lns} with equation \eqref{reduction}, that is, essentially dropping the ``+1" term, you lose asymptotic freedom}. That is, 
\begin{equation}\label{zeta}
   \zeta(s,-1) = \frac{-\Gamma(1-s)}{2\pi i}\int_H dz(-z)^{s-1} \frac{e^{-z}}{1-e^{-z}} 
\end{equation}
which is entire for all \(s\). Under a change of variables \eqref{zeta} is expressed as 
\begin{equation}
 \zeta(s,-1) = \frac{\Gamma(1-s)}{2\pi i}\int_H dz (z)^{s-1} \Big[e^{-z}+ \frac{1}{1-e^{z}}\Big] = (-1)^{-s}+ \zeta(s).
\end{equation}
then
\begin{equation}\label{reduction}
  \zeta(s-1,-1) + \zeta(s,-1) + \zeta(s-1,1) - \zeta(s,1) - \zeta(s-1) = \zeta(s-1).
\end{equation}

\begin{figure}[h!]
\centering
\includegraphics[width=.7\textwidth]{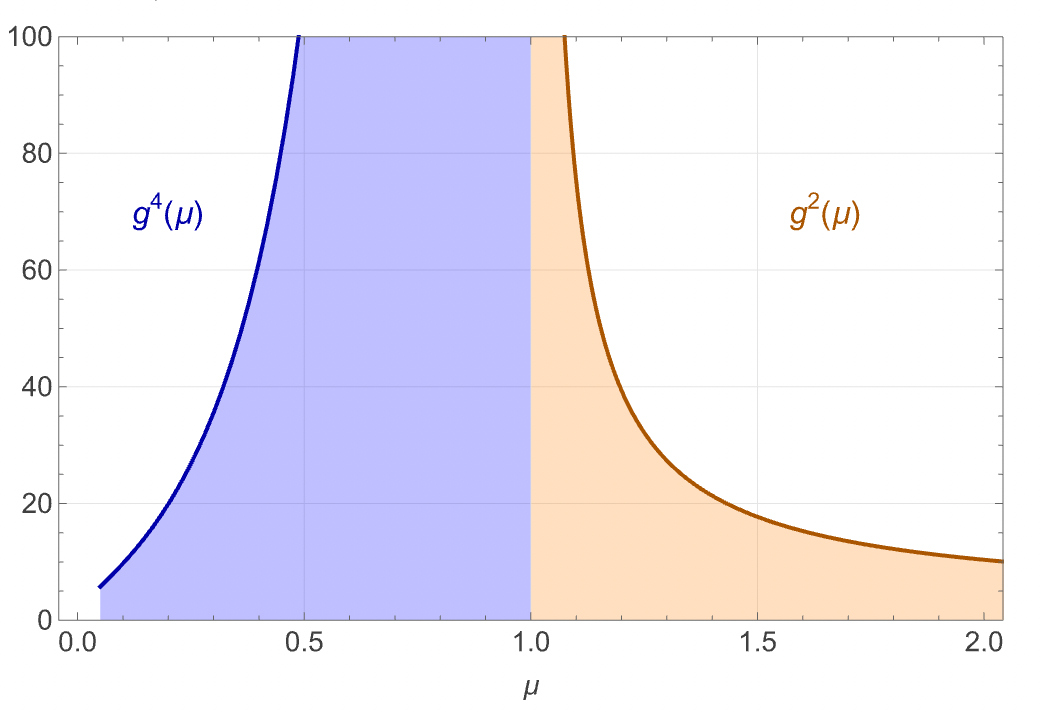}
\caption{\label{complete} \(g^4(\mu)\) and \(g^2(\mu)\) are plotted with \(N=3\) and \(\Lambda=1\) for simplicity.}
\label{both sides}
\end{figure}

Completing the operations on \(s\) in \eqref{lns} gives
\begin{equation}
\begin{split}\label{ln det}
   \frac{\ln(Z)}{\beta V}
   = 2\sum^3_{j=1}-\frac{(B\lambda_j)^2}{g^2(\mu)}-\frac{(B\lambda_j)^2}{48\pi^2}\bigg[11\ln\bigg(\frac{2B|\lambda_j|}{\mu^2}\bigg)-12\zeta'(-1)\bigg]\Bigg|_{\bar{B}_j}
\end{split}
\end{equation}%\sum_j \frac{(B\lambda_j)^2}{4\pi^2} \frac{d}{ds}\Bigg[ \bigg(\frac{2B|\lambda_j|}{\mu^2}\bigg)^{-s}\bigg(\zeta(s-1)+1\bigg) \Bigg]_{s=0}
where the bare coupling has been upgraded to a function of the running scale \(\mu\), and the integral over \(B\) has been approximated via Laplace's method, thus \(\bar{B}_j\) are minimizing values of \eqref{ln det}. The eigenvalues \(\lambda_j\) come in plus-minus pairs, with the inclusion of one zero eigenvalue. For \(N=3\) this eigenstructure represents color and anti color charges. The pressure can be expressed with six degrees of freedom in color space, represented as twice the sum over \(j\) ranging from 1 to 3. Setting \(\frac{d}{d\ln\mu}\ln Z =0\), removes dependence on the running scale \(\mu\) and derives the \(\beta\)-function as
\begin{equation}
    \frac{d g(\mu)}{d \ln (\mu)} = -N\frac{11}{3}\frac{g^3(\mu)}{(4\pi)^2}
\end{equation}
\begin{figure}[h!]
\centering
\includegraphics[width=.8\textwidth]{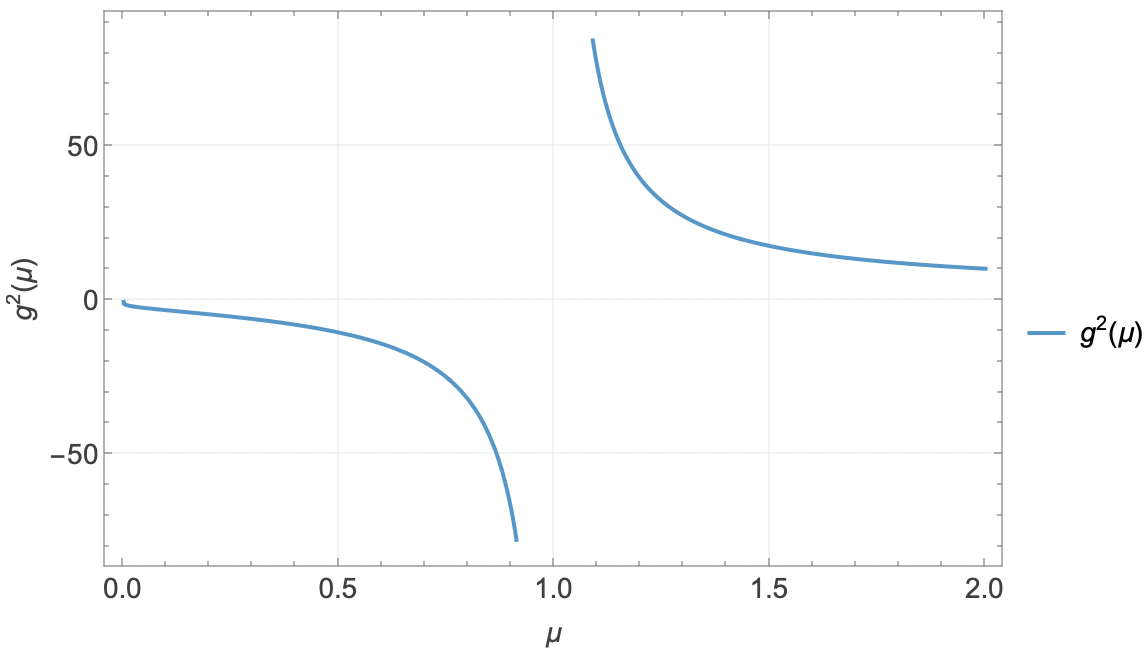}
\caption{ \(g^2(\mu)\) is plotted with \(N=3\) and \(\Lambda=1\) for simplicity.}
\label{run fig}
\end{figure}
matching the one-loop Yang-Mills \(\beta\)-function in \(3+1\)d. The full \(\beta\)-function is plotted in figure \ref{run fig}. Similarly, the running coupling is given as
\begin{equation}
    g^2(\mu) = \frac{48\pi^2}{N11\ln\big(\frac{\mu^2}{\Lambda^2}\big)}.
\end{equation}
 As noted in \cite{romatschke2024life}, the running coupling is defined above and below the Landau-pole scale. The non-perturbative theory is invariant to the running coupling and is therefore well defined for all values of \(g(\mu)\). The traditional picture of the running coupling, as only being above the Landau pole scale, could likewise be used. The realization of the running coupling below the scale of the Landau pole is a mathematically consistent rendering of \(g(\mu)\), and also an empirically falsifiable hypothesis. As \(\mu\) approaches the Landau pole from the UV region \(g(\mu)\) diverges. Because of this, below the Landau pole scale, for a color charge to be unbound, it would have to overcome an infinite potential barrier. Consequently, only color singlets can exist as physical degrees of freedom in the IR, which is shown as the purple region in Figure \ref{both sides}. Thus, when \(g^2(\mu)\) becomes negative, all interactions are mediated exclusively through the exchange of color singlets. This imposes a four-vertex minimum\footnote{It is important to note that it is not the case that the theory is some how different in IR in terms of calculating \(Z\). Because the theory is nonperturbative, \(Z\) is fully invariant to \(g(\mu)\). To be clear, I am not using perturbation theory; instead I am implementing an infrared cutoff of single-gluon exchange, and using diagrams to show the coupling order of singlet interactions. Meaning at lowest order in the IR order all physical process have a \(g^4(\mu)\) effective potential, structurally similar to a Van der waals potential.} for all interactions in the IR, ensuring that all dynamics below the Landau pole scale are of order \(g^{4n}(\mu)\). Only allowing for \(g^{4n}(\mu)\) order interactions in the IR removes the apparent instability generated by negative \(g^2(\mu)\) amplitudes, which would generate colored states, leading to a decay of singlets. 
 This gives a complete picture of the coupling shown in figure \ref{both sides} which possesses UV and IR freedom.

%Incidentally, it seems that the contribution of the \(g\to \infty\) limit from fluctuations vanishes as, as under a commutation of limits\footnote{This is not a rigorous statement, and I do not intend to prove the commutativity of limit operations, although it could be explored in further work.}
%\begin{equation}
   % \lim_{g\to \infty}Z_a =  \int\mathcal{D}a \lim_{g\to \infty}e^{-\int_x\frac{g_(\mu)^2}{4}(f^{abc}a_\mu^b a_\nu^c)^2}.
%\end{equation}
The fully renormalized pressure is now
\begin{equation}
     \frac{\ln(Z)}{\beta V}
   = -2\sum^3_{j=1}\frac{(B\lambda_j)^2}{48\pi^2}\bigg[11\ln\bigg(\frac{2B|\lambda_j|}{\Lambda^2}\bigg)-12\zeta'(-1)\bigg]\Bigg|_{\bar{B}_j}.
\end{equation}
The gap equation for each \(B_j\) is solved by
\begin{equation}
    \bar{B}_j =  \frac{e^{\frac{12}{11}\zeta'(-1)-\frac{1}{2}} \Lambda ^2}{2 \lambda_j }, 0
\end{equation}
giving 
\begin{equation}
    P= N\frac{11}{12} \frac{\Lambda^4}{(4\pi)^2}e^{\frac{24}{11}\zeta'(-1)-1} \approx 15 N\times \bigg(\frac{\Lambda}{10}\bigg)^4.
\end{equation}
\begin{figure}[h!]
\centering
\includegraphics[width=.8\textwidth]{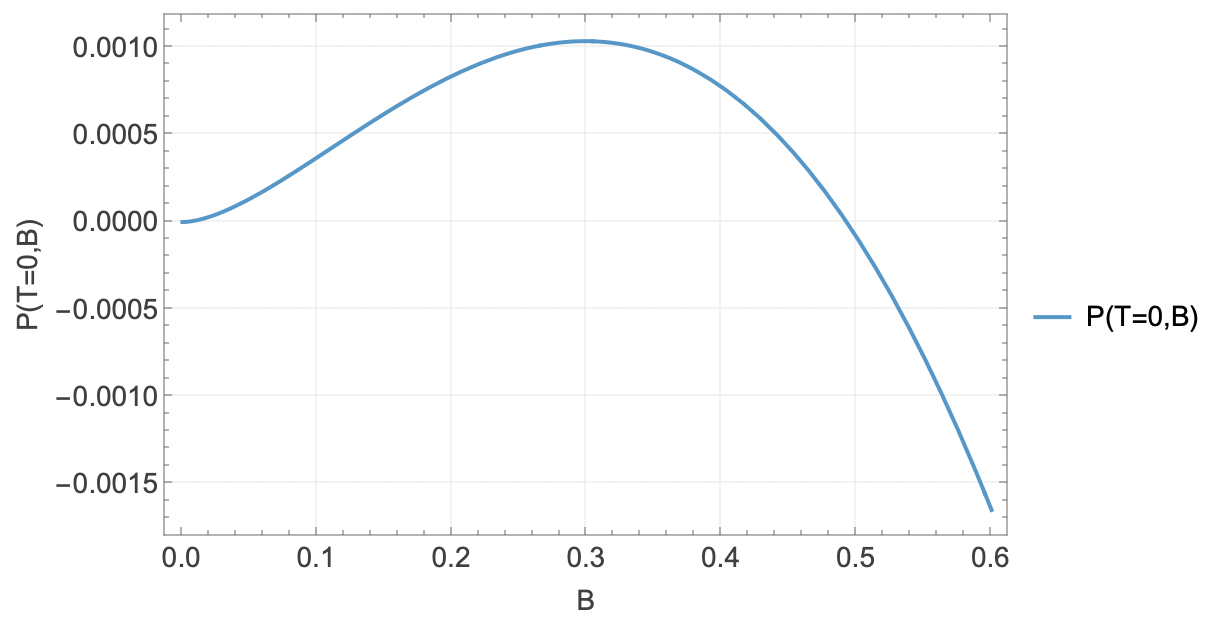}
\caption{\label{graph} P(T=0, B) for a single \(\lambda\) is plotted with \(\Lambda=1\) and \(\lambda=1\) for simplicity.}
\label{pressure fig}
\end{figure}
For \(N=3\) and \(\Lambda\approx 300\) MeV the pressure evaluates to \(P\approx 3.6 \times 10^7\textbf{ }\text{MeV}^4\), and the fourth root of the pressure, associated with the bag constant in QCD, is \(P^{1/4} \approx 77.6 \) MeV. The pressure in pascals is \(P\approx 7.5\times 10^{32}\) Pa, giving a pressure in the range of those found in the inner crust of neutron stars. 

\section{Conclusion}
\paragraph{}
In this work I have identified the zero mode of a covariantly constant and self-dual semiclassical Yang-Mills expansion as resulting from a gauge redundancy of the background field. An additional bosonic gauge fixing term was added in light of this. Then with the use of zeta functions the one-loop path integral was regulated, giving both a zero and non-zero solution to the gap equation, and the correct \(\beta\)-function. The complete picture of the \(\beta\)-function was illustrated by only allowing for interactions that go like multiples of \(g^4(\mu)\) at scales below the Landau pole due to color confinement. This work can be extended to include chiral or massive fermionic fields and finite chemical potentials.

\bibliographystyle{unsrt}
\bibliography{bibliography.bib}
\end{spacing}
\end{document}